\PassOptionsToPackage{dvipsnames}{xcolor}
\documentclass[10pt,sigconf,nonacm,screen]{acmart}

\usepackage[all]{nowidow}
\usepackage{xcolor}
\usepackage{subcaption}
\usepackage{hyperref}
\usepackage{acronym}
\usepackage{siunitx}
\usepackage{pdfpages}

\usepackage[capitalise]{cleveref}
\crefformat{section}{\S#2#1#3} 
\crefformat{subsection}{\S#2#1#3}
\crefformat{subsubsection}{\S#2#1#3}

\begin{document}
\includepdf[pages=-, scale=1.0]{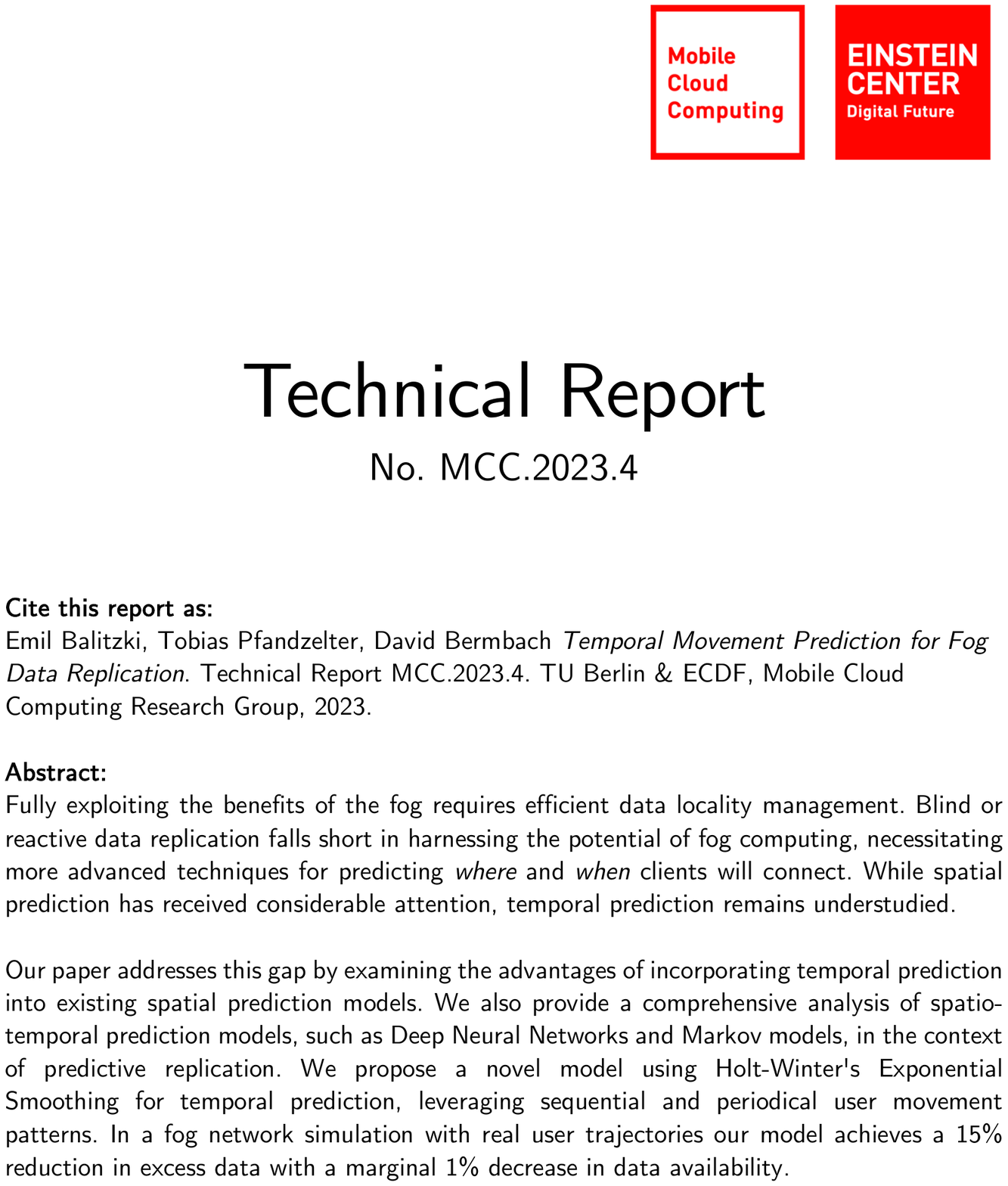}

\author{Emil Balitzki}
\affiliation{%
    \institution{TU Berlin \& ECDF}
    \city{Berlin}
    \country{Germany}}
\email{emb@mcc.tu-berlin.de}

\author{Tobias Pfandzelter}
\affiliation{%
    \institution{TU Berlin \& ECDF}
    \city{Berlin}
    \country{Germany}}
\email{tp@mcc.tu-berlin.de}

\author{David Bermbach}
\affiliation{%
    \institution{TU Berlin \& ECDF}
    \city{Berlin}
    \country{Germany}}
\email{db@mcc.tu-berlin.de}

\title{Predicting Temporal Aspects of Movement for Predictive Replication in Fog Environments}

\begin{abstract}
    To fully exploit the benefits of the fog environment, efficient management of data locality is crucial.
    Blind or reactive data replication falls short in harnessing the potential of fog computing, necessitating more advanced techniques for predicting \textit{where} and \textit{when} clients will connect.
    While spatial prediction has received considerable attention, temporal prediction remains understudied.

    Our paper addresses this gap by examining the advantages of incorporating temporal prediction into existing spatial prediction models.
    We also provide a comprehensive analysis of spatio-temporal prediction models, such as Deep Neural Networks and Markov models, in the context of predictive replication.
    We propose a novel model using Holt-Winter's Exponential Smoothing for temporal prediction, leveraging sequential and periodical user movement patterns.
    In a fog network simulation with real user trajectories our model achieves a 15\% reduction in excess data with a marginal 1\% decrease in data availability.
\end{abstract}

\maketitle

\section{Introduction}
\label{sec:introduction}

By integrating regional data storage and servers with a cloud environment, fog computing offers advantages such as low latency and reduced bandwidth requirements~\cite{paper_bermbach2017_fog_vision}.
As application clients constantly move and connect with different fog nodes~\cite{TOB}, maintaining quality of service (QoS) becomes challenging and reactively transferring data becomes impractical.
Predicting \textit{where} (spatial) and \textit{when} (temporal) data will be needed can significantly reduce latency and enhance overall QoS~\cite{PER}.

In order to predict future location user location is compared with previous trajectory data.
Spatial prediction~\cite{EVA,TEMP,22SUR,21SUR, COMP,SAT,SYS} can benefit from complementary contextual information to improve prediction accuracy~\cite{COMP,22SUR,HD,TAT,SAT,SYS}, e.g., topical preferences~\cite{PROB}, social group influence~\cite{WHERE}, or traveling group influence~\cite{PER}.
Adding temporal context can affect predictive replication accuracy and could be combined with almost any prediction model~\cite{COMP,22SUR,TAT,TEMP,PROB}.
Temporal prediction could decrease data replica holding time in a fog environment but has not been studied extensively~\cite{TOB}.

Between available models and characteristics of a fog data replication use case, a compromise must be made between model applicability to a specific use case and its ability to mine spatio-temporal patterns.
Therefore, in this paper, we make the following contributions:

\begin{itemize}
    \item We compare and classify spatio-temporal prediction models and discuss their characteristics and applicability for predictive replication in a fog environment problem (\cref{sec:prediction_models}).
    \item We discuss the characteristics of the temporal context in the context of predictive replication (\cref{sec:temporal_prediction}).
    \item We propose a Temporal Fusion Multi Order Markov Model (T-FOMM) with different extensions for temporal prediction (\cref{sec:t-fomm}).
    \item We evaluate our model in a simulation of a fog environment using real user movement data (\cref{sec:evaluation}).
\end{itemize}

\section{Background}
\label{sec:background}

We briefly summarize concepts of fog computing and predictive replication.

\subsubsection*{Fog computing}
\label{sec:fog_computing}

\begin{figure}
    \centering
    \includegraphics[width=\linewidth]{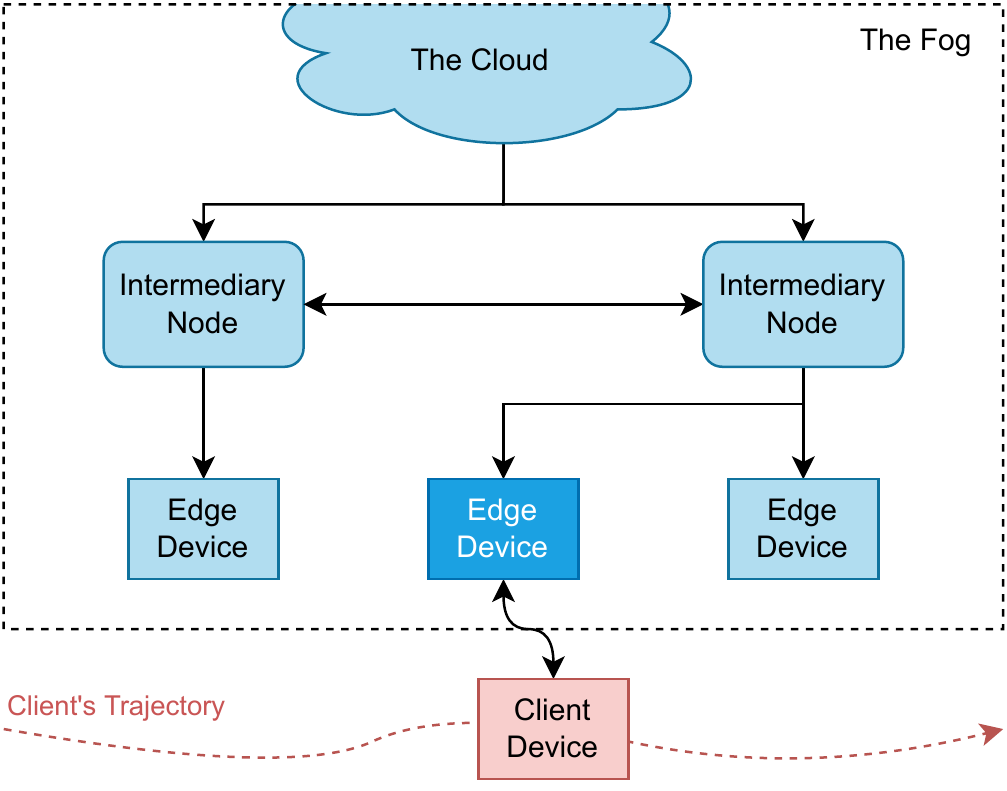}
    \caption{A fog architecture with a moving client connected to the closest edge device~\cite{paper_bermbach2017_fog_vision}.}
    \label{fig:fog}
\end{figure}

Fog computing is a combination of a highly scalable cloud, edge devices, and intermediary nodes, as shown in \cref{fig:fog}.
While centralization in cloud computing has cost benefits, ease-of-use, elastic scalability, and the illusion of infinite resources~\cite{bonomi2012fog,paper_bermbach2017_fog_vision}, data centers are commonly placed far from the end users, resulting in increased latency for data manipulation.
In fog computing, data is stored decentralized, closer to the user.
This change makes the data transfer from the storage to the processing device faster and more efficient while allowing the long-term analysis to still occur in the cloud.
Additionally, due to the data being local, the network usage is decreased, requiring fewer bandwidth resources.
In addition to the cloud, the fog consists of numerous, usually low-spec hardware, heterogeneous IoT devices, with limited computations in terms of processing power and storage~\cite{bonomi2012fog}.
Such devices generate enormous amounts of data, which is stored and processed in the intermediary fog nodes near the edge.

In order to utilize the low latency characteristic of the fog, user-specific data needs to be kept near the user.
This can be cumbersome when application clients constantly movie through the physical world, connecting to different fog nodes~\cite{TOB}.
In an effort to solve this problem, the data can be replicated to multiple fog nodes~\cite{pfandzelter2023fred,poster_hasenburg2020_towards_fbase,techreport_hasenburg2019_fbase}.
However, choosing the nodes to which to replicate the data to is a complex problem, thus in order to optimize the resources, knowing \textit{where} and \textit{when} the end user will access the data is required.

\subsubsection*{Predictive Replication}
\label{sec:predictive_replication}

\begin{figure}
    \centering
    \begin{subfigure}[b]{0.32\linewidth}
        \centering
        \includegraphics[width=\linewidth]{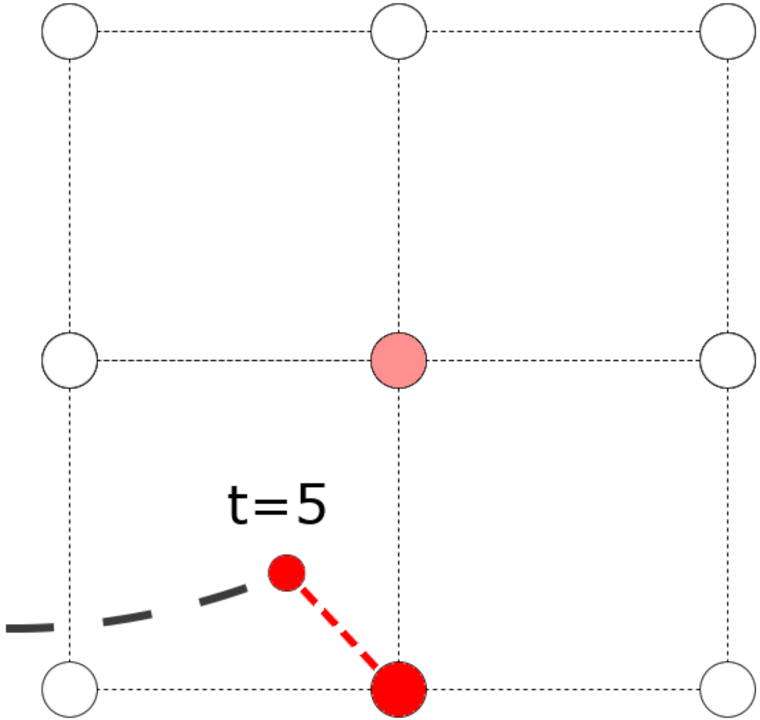}
        \caption{Initial fog node connection}
        \label{fig:pr1}
    \end{subfigure}
    \hfill
    \begin{subfigure}[b]{0.32\linewidth}
        \centering
        \includegraphics[width=\linewidth]{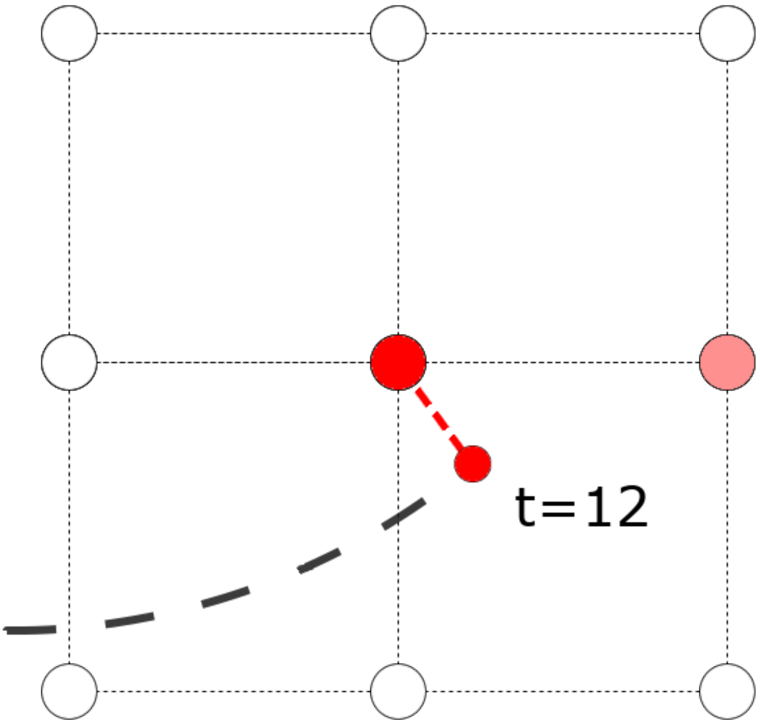}
        \caption{Change of connected node}
        \label{fig:pr2}
    \end{subfigure}
    \hfill
    \begin{subfigure}[b]{0.32\linewidth}
        \centering
        \includegraphics[width=\linewidth]{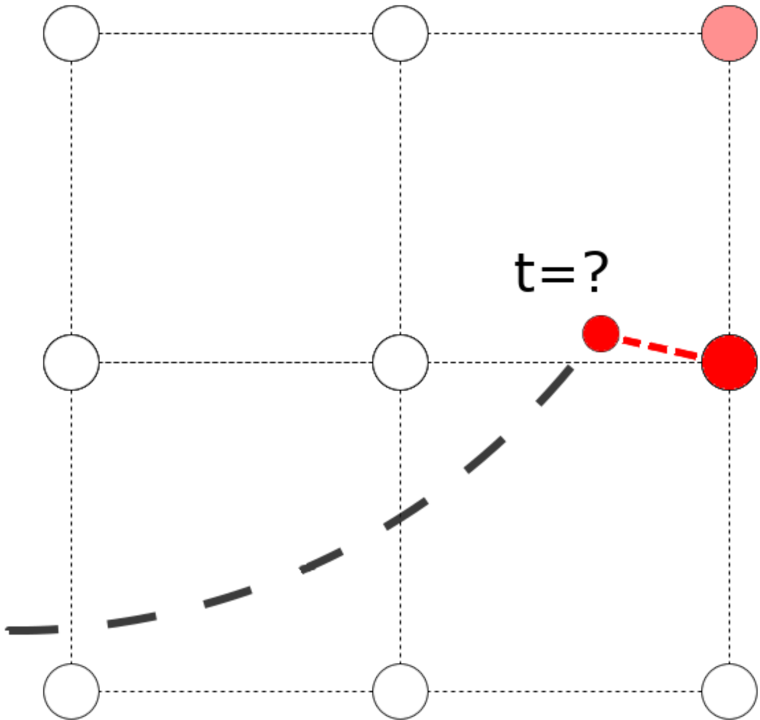}
        \caption{Future fog node connection}
        \label{fig:pr3}
    \end{subfigure}
    \caption{Example spatio-temporal predictive replication in the fog where a client moves through the fog.
        A spatial prediction has to be made about the future node to which the data should be replicated to (bright red node).
        Temporal prediction is necessary to know when movement will occur.}
    \label{fig:pr}
\end{figure}

Predictive replication can be used to proactively replicate the data to the fog node before it will be accessed by the user~\cite{TOB}.
An example for a spatio-temporal Predictive Replication scenario can be seen in \cref{fig:pr}:
First, a spatial prediction is made where the predictor predicts the next replica location(s) based on previous trajectory data.
Second, in order to decrease the excess data, the predictor needs to know \textit{when} such replication should happen, thus adding temporal context to the prediction.
With an introduction of the temporal context, the main quality to be predicted is the duration of the stay at the current location.
Replicating the data immediately after arrival would increase the excess data, as a user might stay longer at a current node than it is needed for replicating the file~\cite{TOB}.
Thus, in order to compensate for this, a value composed of the estimated transfer time, optional buffer, and the sojourn time can be calculated~\cite{TOB,COMP}.
This will assure that the file will be copied just in time for it to be accessed and not stay on the node longer than necessary.

In predictive replication in a fog environment, the client-sided data does not need to be stored in the exact node which will be predicted~\cite{TOB}.
This may increase excess data but decrease access latency in a case of a failed prediction due to the data being available in several highly probable next nodes.
Further, as the fog network may change dynamically, the prediction model has to be easily expandable to compensate.

\section{Spatio-Temporal Prediction}
\label{sec:spatio_temporal_prediction}

In this section, we summarize the characteristics, benefits, and limitations of different location prediction models.
We further discuss how the characteristics of predictive replication and the addition of the temporal context influence the choice of the prediction model, giving a prediction models' classification based on the correlation between the spatial and temporal predictors.
Furthermore, we justify our spatio-temporal prediction model choice and describe it in more detail.

\subsection{Prediction Models}
\label{sec:prediction_models}

While there exist many distinct machine learning based prediction models~\cite{HYB,PHD}, e.g., clustering techniques~\cite{10.1145/2370216.2370421,ying2014mining,alvarez2010trip}, Bayesian models~\cite{COMP,BAY, ye2011exploiting}, neural networks~\cite{10.1145/1280940.1280982,petzold2005next}, or decision trees~\cite{MOB}, the most widely used are the state-based techniques, mainly the \emph{LeZi} family~\cite{5545643}, pattern matching algorithms~\cite{pulliyakode2014modified,10.1093/comjnl/40.2_and_3.67} and Markov models~\cite{TOB,liao2007learning,lee2006modeling,song2006predictability}, including hidden Markov models~\cite{simmons2006learning,letchner2006trip} or Factorizing Personalized Markov Chain~\cite{rendle2010factorizing}.

Due to ease of use, simplicity, efficiency, generality, and domain independence, Markov models have been widely adopted for location predicting~\cite{REN,HYB,zeng2021predicting,WHERE,EVA}.
In Markov models, the notion of states can be directly mapped to the locations, and the transitions between the states to the movement between these locations~\cite{TOB}.
They calculate the probabilities for moving from one state to the other, thus from one location to the other, based on the users' trajectory data.
Markov models are easily expandable, with more locations possible to be added without additional costs~\cite{COMP}, thus performing well in online learning scenarios.
Additionally, Markov models require low modeling effort~\cite{COMP,TOB}, low computing costs~\cite{COMP,bui2017survey}, and show relatively good stability.
Markov models, however, struggle with mining long-term sequences~\cite{22SUR}, not taking the full movement history of the user into account.
They further require a long time for relearning, making many inaccurate predictions during relearning phases~\cite{COMP, REN}.
Additionally, Markov models show poor performance when dealing with irregularity of visits~\cite{22SUR} and predicting the pattern's occurrence for the first time~\cite{COMP,HYB}.
For both spatial and temporal predictions, the discretization of the time is needed~\cite{TOB,REN}, thus requiring splitting the time into two or more time intervals due to the state nature of the model.
With several time intervals selected~\cite{TOB}, in order to receive the final prediction, a weight needs to be assigned for each interval, decreasing the model's stability.
Such discretization will often hide many correlations between time intervals which were not chosen.

Deep Neural Networks (DNN) prove to perform better at spatio-temporal predictions than the standard machine learning approaches~\cite{STF,22SUR,COMP,REN, BTDT,DLS,zeng2021predicting,WHERE,21SUR,STAN}, with Recurrent Neural Networks (RNN) being the most widely used ~\cite{STAN,DLS}.
They characterize with automatic feature representation learning~\cite{DLS}, self-exploring the features' patterns, and automatically learning hierarchical feature representations from the raw, heterogeneous, (un)structured spatio-temporal data~\cite{21SUR}.
Moreover, DNNs have a powerful function approximation ability, theoretically fitting any curve with enough layers and neurons, thus allowing for dealing with nonlinear problems, extracting compound features, and resembling complex functions~\cite{DLS,SPPM}.
On the other hand, DNNs require a considerable modeling effort~\cite{COMP,REN}, exhibit high computation and temporal costs for the training process~\cite{COMP} (although low usage costs for an already trained network), and have problems with long relearning phases~\cite{COMP}.
DNN showed bad stability~\cite{COMP}, where a multitude of parameters can have a significant impact on the final prediction accuracy.
Additionally, DNNs are often described as a black box, with difficulty in understanding their intricate inner workings.
Basic DNN exhibit problems with sequential data mining, therefore to solve this issue RNNs were developed, using historical outputs as the inputs for the next predictions.

RNN can be extended with Gated Recurrent Units (GRU) or Long Short-Term Memory (LSTM)~\cite{21SUR,hochreiter2001gradient, MPP}, to improve long-term spatio-temporal pattern learning~\cite{GEO}.
LSTM transfer not only traditional hidden states (short-term memory) but also additional cell states (long-term memory) between layers.
In GRU relevant information is propagated throughout by just the traditional hidden states~\cite{21SUR}, making GRU less resource-intensive with fewer parameters overall.

Convolutional Neural Networks (CNN) and \emph{GraphCNN} have also been used for mining spatio-temporal patterns in sequential data~\cite{DLS, 21SUR,PHD,DLS, karatzoglou2018convolutional,lv2018t,varshneya2017human}.
Users' trajectories can be represented as a matrix with its two dimensions being the row and column IDs of a spatio-temporal grid field.
Such representations can utilize powerful correlations capturing abilities of CNNs or the GraphCNN to explore node correlations and node features~\cite{DLS}.

\subsection{Temporal Prediction}
\label{sec:temporal_prediction}

The temporal context is mostly influenced by sequential, periodical, or personal preference patterns.
Sequential information focuses on consequent location visits and their correlations based on the order of visits.
For example, it is likely for a shopping center's database to be accessed right after accessing the database located in the office building.
The next location heavily depends on the previous location, making the sequential temporal information one of the most important factors in improving the prediction~\cite{TAT,TEMP}.
Next, with temporal and spatial regularities of people's movement~\cite{HYB} periodical patterns are revealed, characterizing the monthly, weekly, daily, and season-based moving patterns~\cite{TAT,TEMP,SAT,zeng2021predicting,WHAT}.
Strong temporal cyclic patterns for human mobility have been observed, influencing the movement in terms of the hour of the day and day of the week~\cite{TEMP}.
Compared to short-term sequential patterns, periodic patterns can capture more fine-grained temporal visiting behaviors~\cite{TAT}.
Lastly, personal preference patterns describe the user's preference changing over longer periods, leading to different visiting behaviors in different time periods~\cite{TAT}.
These changing preferences can become a problem for some prediction models.

Based on the correlation between the spatial and temporal predictors, we have derived a classification of spatio-temporal prediction models:
A \emph{single prediction model} is a single model that predicts both spatial and temporal characteristics, a \emph{collaborative model} combines a spatial model with a temporal model where the output depends on the spatial model, and an \emph{independent model} combines spatial and temporal models that are independent of each other.

\subsubsection*{Single Prediction Model}
\label{subsubsec:single_pm}

Liu et al.~\cite{ST-RNN} introduce a spatial-temporal RNN for check-in location that separates the spatial and temporal values into discrete bins in order to produce distance and time-specific transition matrices.
Their model is thus able to extract periodical contexts~\cite{STF,STC,TBS}, outperforming the standard RNN.
Kong and Wu~\cite{HST-LSTM} introduce spatio-temporal relations to internal LSTM gates to mitigate data sparsity.
Zhao et al.~\cite{WHERE} equip LSTM with new time and distance gates for POI recommendation, outperforming competing approaches.
Luo et al.~\cite{STAN} propose a spatio-temporal attention network that exploits the spatio-temporal information of all check-ins with self-attention layers along the trajectory.
Zeng et al.~\cite{zeng2021predicting} propose a self-attention RNN model to explore sequence regularity and extract temporal features according to historic trajectory information.
Their model first processes the sparse data using an embedding layer and transforms it into dense potential representation, which is then fed to the RNN to mine complex long-term dependencies, before a self-attention mechanism captures contextual factors.
Further models are discussed in extensive surveys, e.g., by Chekol et al.~\cite{22SUR}.

Despite their high accuracy, DNN-based methods struggle with online learning, require high modelling effort, and are expensive to train~\cite{capka2004mobility,REN}.
This makes them less compatible with predictive replication than, e.g., Markov models, which in turn often show lower accuracy~\cite{STF,22SUR,COMP,REN,BTDT,DLS,zeng2021predicting,WHERE,21SUR,STAN}.
While Qiao et al.~\cite{HYB} have introduced a Markov model that can capture long-term location sequence patterns, they still require time discretization to be used as a single spatio-temporal prediction model.

\subsubsection*{Collaborative Model}
\label{subsubsec:collaborative_pm}

Collaborative models consist of multiple predictors in order to make the final spatio-temporal prediction~\cite{ZHU2019190,kamara2022ensemble,gidofalvi2012and}.
The temporal predictor depends on the spatial predictor and works simultaneously with or after it.
Ali et al.~\cite{DHSTNet} jointly model spatial and temporal correlations with the use of a CNN-LSTM fusion in order to predict traffic flows in every region of a city.
The authors~\cite{DHSTNet2} further extend this with increased prediction speed and accuracy using attention mechanism.
Chen et al.~\cite{DeepJMT} propose \emph{DeepJMT}, a context-aware deep model that mines evidence from social relationships.
DeepJMT is a hybrid three-component model, that combines a hierarchical RNN-based sequential dependency encoder that captures users' spatial and temporal regularities, spatial and periodicity context extractors to extract location semantics and periodicity, and a co-attention-based social and temporal context extractor that integrates the social influence.
When implementing a collaborative model, the characteristics of all extensions have to be taken into account, making development even more complex.

\subsubsection*{Independent Model}
\label{subsubsec:independent_pm}

In an independent model, prediction models are not cooperative, making each sub-model easier to implement.
Both predictions can be executed one after another or simultaneously.
Spatial predictions can be made using any of the mentioned models, while temporal predictions can be added using different prediction techniques or statistical methods, e.g., simple averaging, regression models, or time series forecasting techniques.
Temporal prediction techniques unsuitable for a specific use case are thus incorporated.
The downside is that spatio-temporal correlations are lost~\cite{DeepJMT,DHSTNet,DLS,GEO}.

\section{Temporal Fusion Multi Order Markov Model (T-FOMM)}
\label{sec:t-fomm}

\begin{figure}
    \centering
    \includegraphics[width=\linewidth]{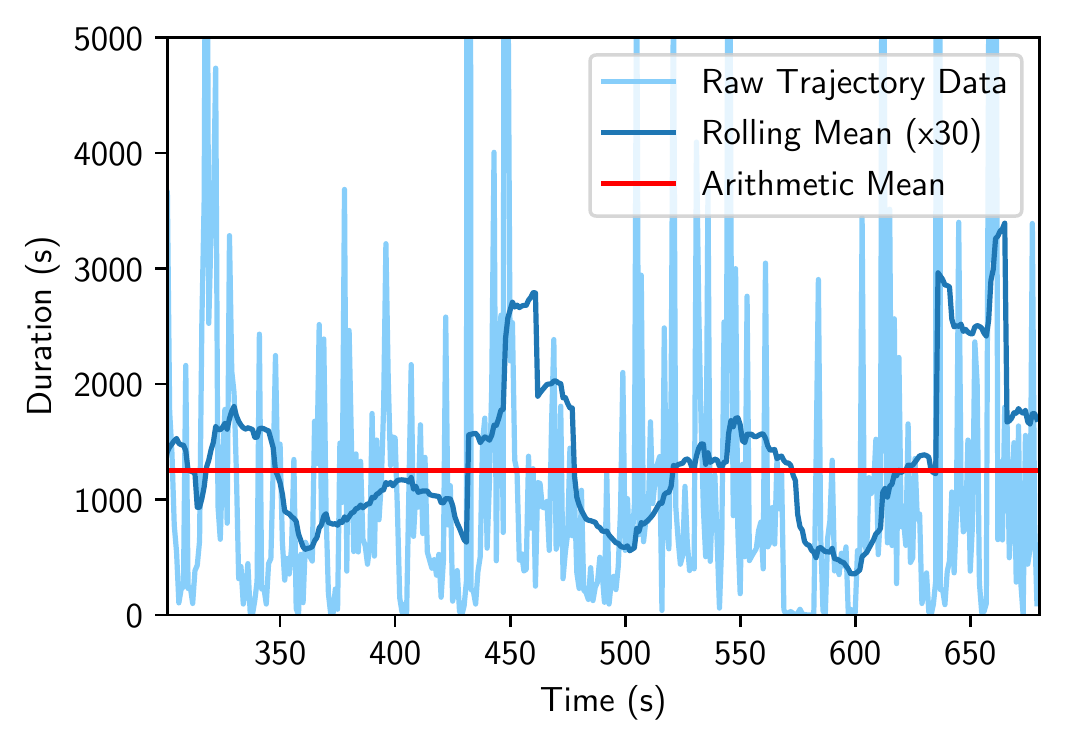}
    \caption{A time series of stay durations from the trajectory data of a single user from the evaluated dataset.
        The \textit{periodical} temporal pattern is illustrated with the rolling mean.}
    \label{fig:sessionality}
\end{figure}

Predictive replication in a fog environment requires online learning and adaptability to changing network topologies.
We propose Temporal Fusion Multi Order Markov Model (T-FOMM), an independent model that combines the Fusion Multi Order Markov Model introduced by Bellmann et al.~\cite{TOB} with an independent temporal model.
FOMM showed to perform well in an expanding and changing network topology and is capable of online learning.
FOMM models spatio-temporal patterns by combining sub-models with different time discretization, e.g, by day, week, or month.
It uses a simple arithmetic mean, which omits periodical patterns as shown in \cref{fig:sessionality}.
Due to T-FOMM being independent prediction model, the temporal prediction is decoupled from the spatial prediction.
With the temporal predictor freely exchangeable, we propose and evaluate several temporal predictors: percentiles (\cref{subsec:t-fomm-pctl}), temporal discretization (\cref{subsec:t-fomm-da}), and Holt-Winter's exponential smoothing (\cref{subsec:t-fomm-hwes}).

\subsubsection*{T-FOMM(PCTL): Percentiles}
\label{subsec:t-fomm-pctl}

Instead of the arithmetic mean, we calculate a $k$\textsuperscript{th} percentile from the set of durations spent at a node.
Compared to the arithmetic mean, this allows for more control of the impact of outliers in the data.

\subsubsection*{T-FOMM(TD): Temporal Discretization}
\label{subsec:t-fomm-da}

Periodical temporal patterns can be taken into account by discretizing stay duration into sets, e.g, \emph{monthly}, \emph{day of the week} and \emph{hourly}, similar to the original FOMM.
For a prediction, we use the arithmetic mean from the corresponding set.
If there is no value for a given set, the predicted duration is approximated from the mean of the closest non-empty sets.

\subsubsection*{T-FOMM(HWES): Holt-Winter's Exponential Smoothing}
\label{subsec:t-fomm-hwes}

\begin{figure}
    \centering
    \includegraphics[width=1\linewidth]{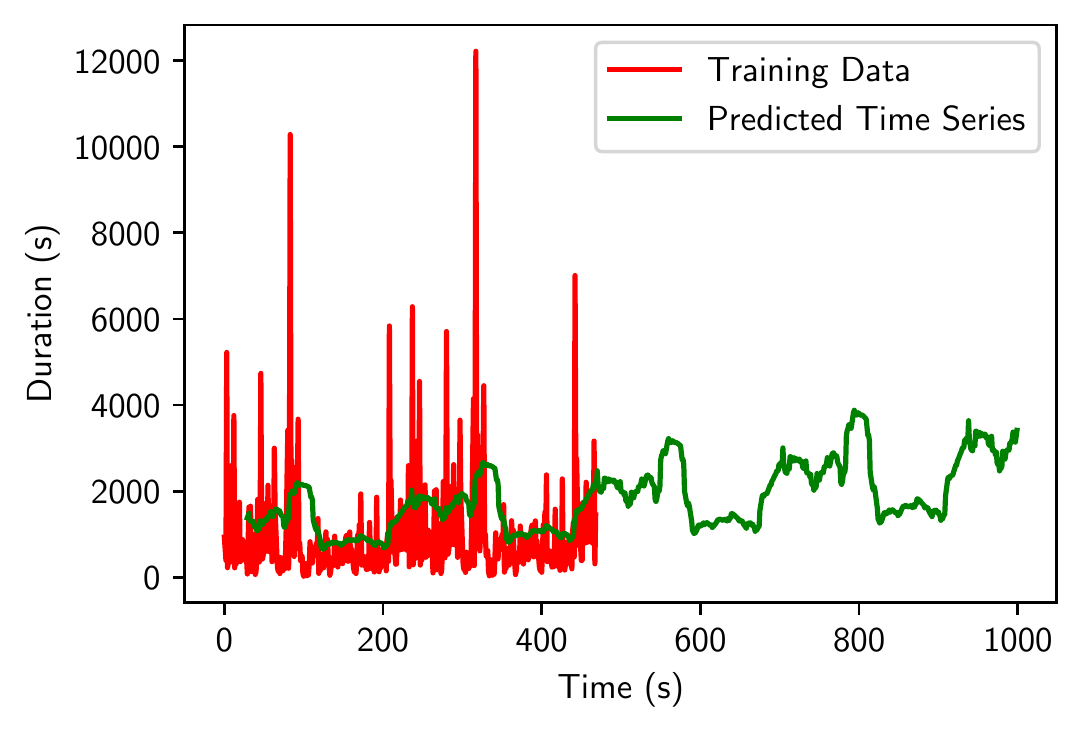}
    \caption{Holt-Winter's Exponential Smoothing time series forecasting smoothed by a moving average of 20, with the user data split for a single user.}
    \label{fig:hwes}
\end{figure}

Holt-Winter exponential smoothing (HWES)~\cite{winters1960forecasting} is an exponential smoothing technique used for time series forecasting, where decaying weighted average of raw data is combined with trend and seasonality to model the periodical and sequential patterns.
With growing training data, we collect data on different levels of the FOMM predictor, namely the \emph{discretization}, \emph{node} or \emph{user} data split.
The discretization level is the most fine-grained option, splitting the training data sets for specific discretization sub-models of the FOMM.
The node level merges data from sub-models for each of the predicted fog nodes.
The user level collects all training data for each user for all possible predicted nodes, creating the biggest training data sets from all data split options.

For a prediction, we calculate a HWES model for all collected data.
We show an example of an HWES prediction for more than one point in \cref{fig:hwes}, with a visible periodical temporal pattern and a subtle overall increasing trend.
In order to compensate for the pause between the last data item and the time of the prediction T-FOMM(HWES) aggregates the predicted values until a duration of the pause is reached.
\section{Evaluation}
\label{sec:evaluation}

We evaluate our proposed T-FOMM models in simulation with moving clients based on location traces and record their connections with a fog network model.
Our simulation environment is available as open-source software.\footnote{\url{https://github.com/OpenFogStack/temporal-location-prediction}}
When a fog node is instructed to replicate a data set for a user, it is downloaded after a specified transfer time of 5 minutes, similar to replica management in \emph{FReD}~\cite{pfandzelter2023fred,poster_hasenburg2020_towards_fbase,techreport_hasenburg2019_fbase}.
All datasets are specific to a user, and users randomly appear and disappear from the network.
For the network model, we use a simple, evenly-spaced grid network of fog nodes with a fixed data transfer time between all nodes.
User trajectories are based on the \emph{GeoLife} GPS trajectory dataset of 182 users in Beijing, China, between 2007 and 2012~\cite{zheng2008understanding,zheng2010geolife,zheng2009mining}.
Moving clients are assumed to connect to the physically closest fog node receiving the best QoS from the replica management application.
We use two main metrics to evaluate prediction methods:
Data availability indicates the percentage of time when a user's closest fog node contains the necessary data (higher is better).
Excess data shows the relative amount of data stored at other fog nodes, i.e., data that is replicated to the wrong location (lower is better).

\subsubsection*{Baseline}

The baseline used to evaluate our temporal prediction extensions is the original FOMM with all extensions~\cite{TOB}.
Further, we compare two naive ideal models:
A \emph{keep-on-closest} model reactively replicates data to the closest fog node of a user (no prediction, no excess data).
An \emph{always-on-all} model always replicates data to all nodes (highest excess data and highest data availability).

Reactively replicating data with the \emph{keep-on-closest} model achieved 61.43\% availability and 0\% excess data.
The \emph{always-on-all} model achieves 99.95\% availability.
We attribute the 0.05\% of time when data is not available to the startup phase, where replicas have to be downloaded to all nodes first.
The cost for this is a 396535.77\% of excess data.
The FOMM model achieves 72.90\% availability and 65.44\% excess data.

\subsubsection*{T-FOMM(PCTL)}

\begin{figure}
    \centering
    \includegraphics[width=\linewidth]{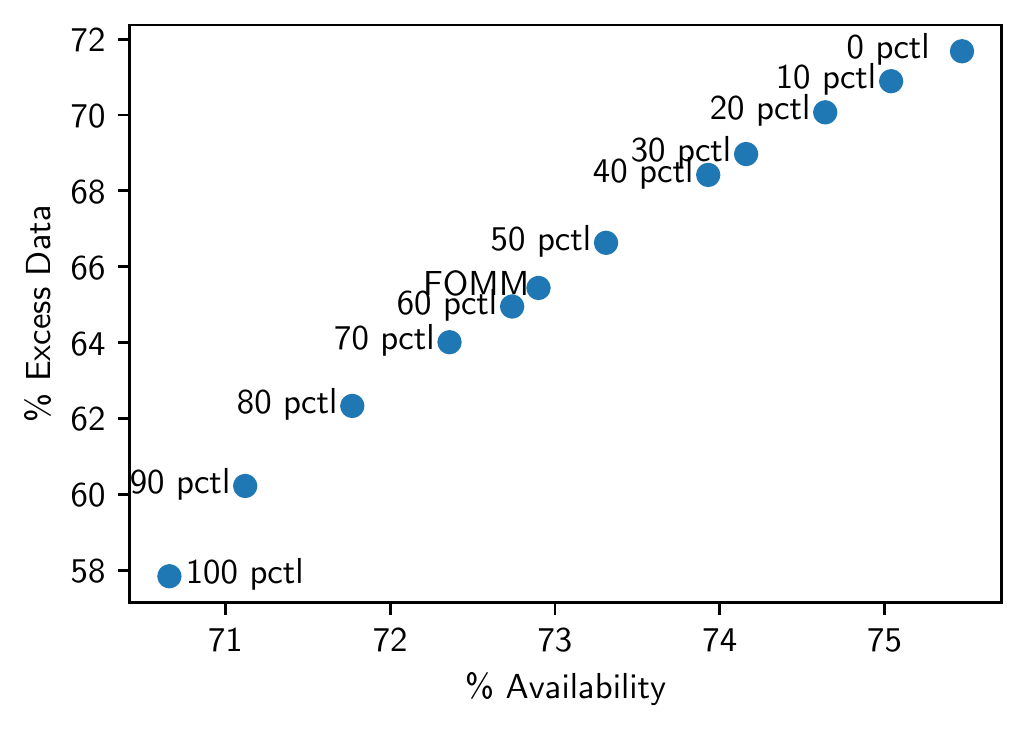}
    \caption{T-FOMM(PCTL) model results show that availability and excess data increase with lower percentiles.
        Baseline FOMM model included for comparison.}
    \label{fig:percentiles}
\end{figure}

We show the results of T-FOMM(PCTL) with different percentiles between the 0\textsuperscript{th} and 100\textsuperscript{th} in \cref{fig:percentiles}.
There is a trade-off between excess data and availability, which both increase with lower percentiles.
The bigger the selected percentile, the higher the possible returned duration, therefore the replication of data happens at later times.
Conversely, replication happens too soon with lower percentiles, increasing excess data.
Nevertheless, selecting percentiles is a direct way to influence this trade-off.

\subsubsection*{T-FOMM(TD)}

\begin{table}
    \centering
    \caption{T-FOMM(TD) Models Evaluation}
    \label{tab:eval_da}
    \begin{tabular}{l r r}
        \toprule
        Discretization Method & Availability & Excess Data \\
        \midrule
        Days of Week          & 73.00\%      & 65.56\%     \\
        Hours                 & 72.94\%      & 65.49\%     \\
        Months                & 72.90\%      & 65.31\%     \\
        Days of Week (median) & 73.24\%      & 66.33\%     \\
        Hours (median)        & 73.17\%      & 66.05\%     \\
        Months (median)       & 73.05\%      & 65.83\%     \\
        \bottomrule
    \end{tabular}
\end{table}

\begin{figure}
    \centering
    \includegraphics[width=\linewidth]{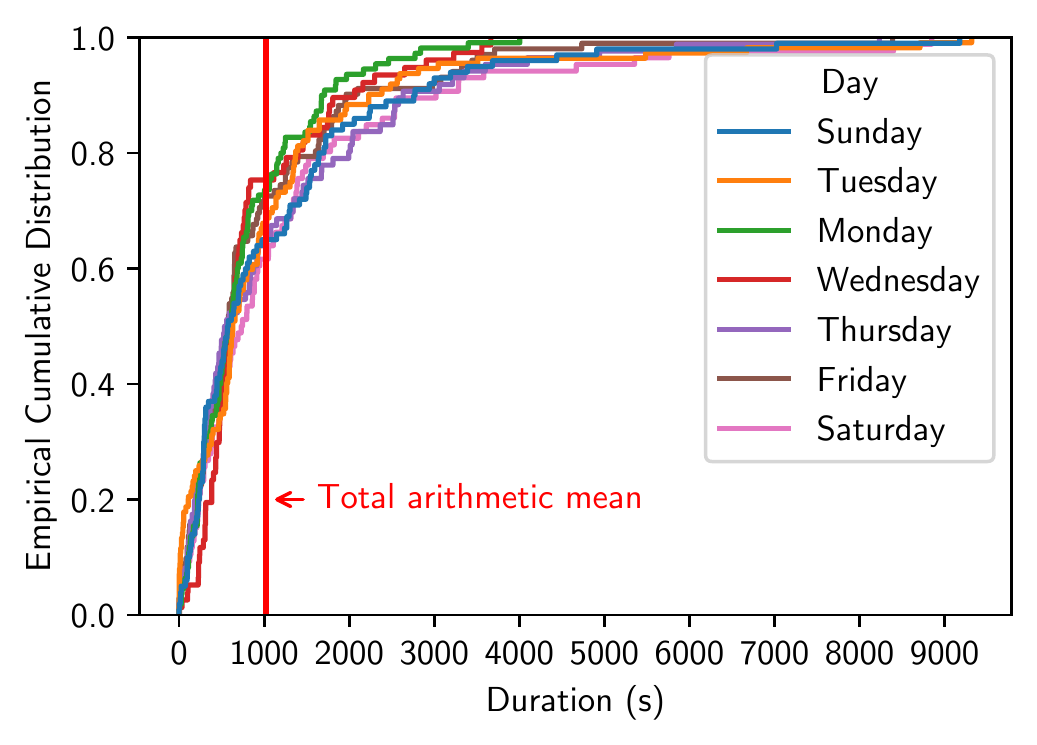}
    \caption{Empirical cumulative distribution for the \textit{days in a week} discretization method from a single user's trajectory data.
        With the data cut over the 99\textsuperscript{th} percentile and the red vertical line showing the arithmetic mean over cumulative values from all days.}
    \label{fig:ecdf}
\end{figure}

We evaluate the T-FOMM(TD) with three independent temporal sets: hours in a day, days in the week, and months.
As shown in \cref{tab:eval_da}, the T-FOMM(TD) achieve a negligible improvement of at most 0.1\% over the baseline FOMM, while generally exhibiting higher excess data.
Those results may be a result of the model having sparser data for specific discretization sets, thus generally calculating lower averages for each prediction.
In turn, quicker replication thus increases excess data.
The overall negligible improvements show, that for the sparse data standard arithmetic mean from the cumulative set is good enough for the temporal prediction, thus showing no need for the discretization of the values.
A possible improvement could be achieved with more available data.
We also evaluate the T-FOMM(TD) models with a median instead of simple arithmetic mean.
The medians for each of the discretized days result in durations around 600s, while the total arithmetic mean equals roughly 1000s, as shown in \cref{fig:ecdf}.
The lower durations reflect higher availability and excess data due to the replication happening faster.

\subsubsection*{T-FOMM(HWES)}

\begin{table}
    \centering
    \caption{T-FOMM(HWES) Models Evaluation}
    \label{tab:eval_hwes}
    \begin{tabular}{l r r}
        \toprule
        Data split     & Availability & Excess Data \\
        \midrule
        Discretization & 72.57\%      & 63.15\%     \\
        Node           & 72.84\%      & 61.58\%     \\
        User           & 71.87\%      & 49.61\%     \\
        \bottomrule
    \end{tabular}
\end{table}

We show results for different split levels for T-FOMM(HWES) in \cref{tab:eval_hwes}.
HWES models achieve better excess data as a trade-off of decreased availability, with the T-FOMM (HWES, user) showing a 15.83\% excess data improvement with just 1\% loss of data availability.
This shows superiority of time series forecasting over simpler metrics, with better results as more training data is available.
HWES correctly predicted the trend and seasonality of user mobility, taking into account both sequential and periodical temporal trajectory patterns.

\subsubsection*{Comparison}

\begin{figure}
    \centering
    \includegraphics[width=\linewidth]{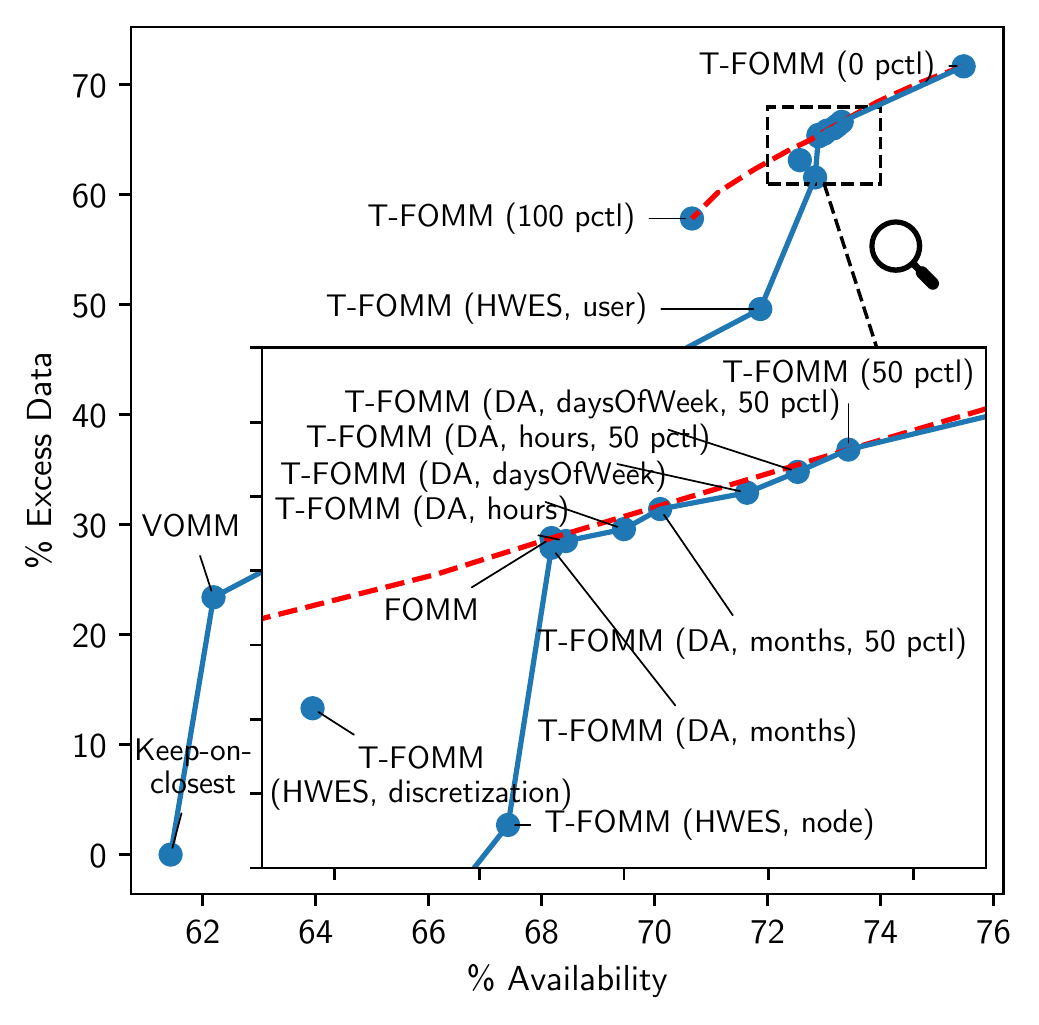}
    \caption{Comparison of models used in our evaluation.
        The blue line represents the Pareto Front. The red dashed line represents the percentiles models. \emph{Keep-on-all} model omitted for scale.}
    \label{fig:pareto}
\end{figure}

There is a clear trade-off between data availability and excess data, which we show in \cref{fig:pareto}.
The Pareto front includes most of the proposed T-FOMM, suggesting they are a valid proposition for selecting a final model.
The standard FOMM model is dominated by the slightly better T-FOMM(TD).
For each model there exists a finer trade-off along the potential percentile line.

\section{Discussion}
\label{sec:discussion}

\subsubsection*{DNN Single Model}

The proposed independent models may miss the spatio-temporal correlations.
A DNN-based approach may improve this, yet we found existing solutions incompatible with predictive replication in the fog given data sparsity and demand for online adaptability.
Advances such as \emph{lifelong learning} have shown the possibility of online training of DNNs when enough training data is available~\cite{castro2018end,li2017learning,van2019three,parisi2019continual,gupta2021continual}.
In order to take advantage of the superiority of GRU or LSTM a different suitable scenario could be considered in future work.
Finally, the use of the attention mechanism~\cite{DLS,zeng2021predicting,GEO,21SUR,GEO} or other state-of-art DNN improvements should be examined.

\subsubsection*{Other Models}

Clustering techniques~\cite{10.1145/2370216.2370421,ying2014mining,alvarez2010trip}, Bayesian models~\cite{COMP,BAY,ye2011exploiting}, or decision trees~\cite{MOB} are possible alternatives for prediction methods.
However, research has not shown them to be more effective that the methods discussed in this paper.
These findings could be quantitatively evaluated in an extensive comparison study in the context of predictive replication in fog computing.

\subsubsection*{Additional Contextual Parameters}

We focus on adding temporal context to spatial prediction in this paper, yet other contextual parameters may also be investigated in the future.
Among them, users' topical preferences~\cite{PROB}, social group influence~\cite{WHERE}, or traveling group influence~\cite{PER} could be taken into account in a prediction model.

\subsubsection*{Evaluation Scenario}

Our evaluation uses a simple artificial grid network of fog nodes, which may not necessarily be representative of real fog infrastructure.
Similarly, the replication duration of five minutes can also be unrealistic in some use cases.
A more realistic complex network consisting of fog nodes and cloud servers connected by links with limited bandwidth with a more realistic data transfer time could change evaluation results.
However, fog networks have yet to be widely available to derive realistic topologies~\cite{rausch2020synthesizing,pfandzelter2023fred}.

\section{Related Work}
\label{sec:relatedwork}

Predicting next locations is applicable to a wide range of fields~\cite{STF, zeng2021predicting, NMDC}, e.g., recommendation systems~\cite{casino2015k}, healthcare and disease transmission control~\cite{solanas2014smart, eubank2004modelling, ajelli2017estimating}, urban sensing and planing~\cite{becker2011tale, reades2007cellular, altomare2016trajectory}, carpooling~\cite{elbery2016proactive}, and sociology~\cite{chittaranjan2013mining, eagle2009inferring}.
Each field has specific requirements to a location prediction approach.
Petzold et al.~\cite{COMP} evaluate several next location prediction methods, focusing on the movement of people in an office building.

Temporal dimensions have received less attention in existing research.
Gao et al.~\cite{TEMP} propose a general framework for exploiting and modeling temporal cyclic patterns in combination with spatial and social data.
They show that user mobility behavior is affected by various temporal patterns that can be modeled as Gaussian mixture distributions.
Zhao et al.~\cite{TAT} propose a time-aware trajectory embedding model.
Zeng et al.~\cite{zeng2021predicting} introduce a DNN-based method with a self-attention mechanism to predict the next location, focusing on temporal features for the final prediction.
Chon et al.~\cite{EVA} explore fine-grained and continuous mobility data for evaluating mobility models.
They find that stay duration is closely correlated to the arrival time at the current location and the return tendency to the next location rather than the sequence of recent locations.
The discussed works, however, do not focus on predictive replication in a fog environment.

Predictive fog data replication has been attempted by Bellmann et al.~\cite{poster_pfandzelter2021_predictive_replica_placement_poster,TOB}, who introduce the FOMM model used as a baseline in our evaluation.
Again, this model focuses on spatial prediction and does not take temporal prediction into account.
Gossa et al.~\cite{gossa2008proactive} propose \emph{FReDI}, a flexible management system for proactive replica placement over a network of proxy-caches.
In contrast to fog data, this assumes a global dataset for all users.
Ara{\'u}jo et al.~\cite{araujo2020cmfog} suggest other Markov model implementations for predictive content migration, focusing again on spatial prediction.
Torabi et al.~\cite{torabi2022data} provide a comprehensive systematic review of current data replica placement approaches.
Gid{\'o}falvi and Dong~\cite{gidofalvi2012and} propose a Markov model single prediction model using time discretization.
Their evaluation does not directly focus on predictive data replication and does not take excess data into account.
Salaht et al.~\cite{salaht2020overview} survey service placement techniques, which have similar requirements as data replication~\cite{paper_bermbach2020_auctions4function_placement,paper_bermbach2021_auctionwhisk,fahs2020voila}.
They conclude that most existing techniques are reactive rather than proactive.

Hasenburg et al.~\cite{techreport_hasenburg2019_fbase,poster_hasenburg2020_towards_fbase,pfandzelter2023fred} discuss the need for replication services for data-intensive fog applications and propose the \textit{FBase} replication service.
Their focus is on providing the software building blocks to enable data replication in the fog, providing APIs for replica location prediction.

\section{Conclusion}
\label{sec:conclusion}

Predictive data replication in the fog can improve data availability for users without leading to excessive replication.
While spatial prediction has been investigated, temporal contexts have largely been ignored in this area in the past.
We have proposed the Temporal Fusion Multi Order Markov Model (T-FOMM) independent prediction model which takes into account the temporal aspect of the prediction.
In evaluation in a simulation, we have shown an improvement of excess data over the baseline without significant decrease in data availability.
By using Holt-Winter's exponential smoothing for time series forecasting we have separated the trend and seasonality of user mobility data, allowing for a more accurate stay duration prediction.

\begin{acks}
    Supported by the \grantsponsor{DFG}{Deutsche Forschungsgemeinschaft (DFG, German Research Foundation)}{https://www.dfg.de/en/} -- \grantnum{DFG}{415899119}.
\end{acks}

\balance

\bibliographystyle{ACM-Reference-Format}
\bibliography{bibliography.bib}

\end{document}